\documentclass[12pt]{article}
\usepackage{mathrsfs}
\usepackage{amsmath}
\usepackage{mathrsfs}
\usepackage{amssymb}
\usepackage{color}
\usepackage{psfrag}
\newcommand{\bea}{\begin{eqnarray}}
\newcommand{\eea}{\end{eqnarray}}
\newcommand{\be}{\begin{equation}}
\newcommand{\ee}{\end{equation}}
\newcommand{\vs}[1]{\vspace{#1 mm}}

\renewcommand{\a}{\alpha}
\renewcommand{\b}{\beta}

\renewcommand{\d}{\delta}

\newcommand{\dsl}{\pa \kern-0.5em /}

\newcommand{\half}{\frac{1}{2}}
\newcommand{\pa}{\partial}

\newcommand{\nn}{\nonumber\\}

\begin{document}
\topmargin 0mm
\oddsidemargin 0mm

\begin{flushright}

USTC-ICTS-17-03\\

\end{flushright}

\vspace{2mm}

\begin{center}

{\Large \bf On 1/4 BPS ((F, D1), (NS5, D5)) bound states of type IIB string theory}

\vs{10}

{\large Qiang Jia$^a$, J. X. Lu$^a$ and Shibaji Roy$^b$}

\vspace{4mm}

{\em

 $^a$ Interdisciplinary Center for Theoretical Study\\
 University of Science and Technology of China, Hefei, Anhui
 230026, China\\
 
 \vs{4}
 
  $^b$ Saha Institute of Nuclear Physics,
 1/AF Bidhannagar, Calcutta-700 064, India\\

\vs{4}

{\rm and}

\vs{4}

Homi Bhabha National Institute\\
Training School Complex, Anushakti Nagar, Mumbai 400085, India}
\end{center}

\vs{10}

\begin{abstract}
We construct two new SL(2,Z) invariant vacua of type IIB string theory which are bound states of 
$(p,q)$ strings with $(m,n)$ 5-branes, written as ((F, D1), (NS5, D5)) and preserve 1/4 of the full space-time 
supersymmetries. For the first case, the strings live inside the 5-brane world-volume and in the
second case the strings are perpendicular to the 5-brane world-volume. In the first case, naively 
one would expect an attractive interaction between the strings and the 5-branes due to attractive
force between F and D5 and also between D1 and NS5. We find that 1/4 BPS bound state exists only
when the vacuum moduli satisfy certain condition which is found to be consistent with the no-force
condition between the branes. No such complication arises for the second case.    
The tension formulae and the various other descendant states which can be obtained by the application
of T-duality for both these bound states are discussed.  
\end{abstract}

\newpage

\section{Introduction}
Type IIB superstring theory in the low energy limit is well-known to possess a so-called Cremmer-Julia or classical  
U-duality SL(2,R) symmetry. Only the discrete subgroup, SL(2,Z) of this, is believed to survive quantum mechanically. 
Therefore, one would expect various vacuum-like BPS states of SL(2,Z) multiplets to exist in this theory.  
The classical U-duality group 
SL(2,R) is useful for the purpose of constructing such SL(2,Z) multiplets. It was Schwarz who first 
explicitly constructed the  SL(2,Z) multiplet of $(p, q)$ strings in this theory \cite{Schwarz:1995dk}. Two of us 
then constructed the SL(2,Z) multiplet of $(m, n)$ 5-branes \cite{Lu:1998vh} and subsequently constructed the 
more complicated non-threshold bound state ((F, D1), (NS5, D5)), where,  (NS5, D5) denotes the $(m, n)$ 
5-branes \cite{Lu:1999uc} and (F, D1) stands for the $(p, q)$ strings living along one of the $(m,n)$ 
5-brane spatial directions while delocalized along the other four. 

However, all the SL(2,Z) multiplets mentioned so far are 1/2 BPS 
non-threshold bound states and their existence lends support to the conjectured quantum 
SL(2,Z) symmetry of type IIB superstring theory. For the 1/2 BPS non-threshold bound state 
((F, D1), (NS5, D5)), it was shown that the quantized charges for its $(p, q)$ strings and $(m, n)$ 5-branes 
can not be completely arbitrary but must satisfy a relation given by,
$(p, q) = k (a, b),\,\, (m, n) = k' (-b, a)$ with $(a, b)$ and $(k, k')$ being two pairs of co-prime integers.
In particular, from this state we can recover 1/2 BPS (F, D5) state by setting $a=1$ and $b=0$ and (D1, NS5) state by
setting $a=0$ and $b=1$. We can also recover a delocalized $(p,q)$ string by putting $k'=0$ as well as $(m,n)$
5-branes by putting $k=0$. However, because of the charge relation we just mentioned we can not recover
either (D1, D5) or (F, NS5) bound state from the general state. This is entirely consistent with the fact that both
these latter states are 1/4 BPS threshold bound states and therefore should not be obtained from 1/2 BPS non-threshold bound state
((F, D1), (NS5, D5)). Also it is known that both the SL(2,R) and the SL(2, Z) preserve the underlying supersymmetry
of a given state on which they act.

It is then natural to expect a new 1/4 BPS ((F, D1), (NS5, D5)) bound state to exist given the existence of 1/4 BPS (D1, D5) and (F, NS5). 
However, we know that there is no-force between D1 and D5 in the threshold bound state (D1, D5)  (or between F and NS5 in (F, NS5)) and so 
if 1/4 BPS ((F, D1), 
(NS5, D5)) bound state indeed exists, we expect that there should also be no-force between the non-threshold (F, D1) and the non-threshold (NS5, D5) 
and this bound state should also be threshold with respect to these two constituent non-threshold bound states.  While it is true that there is no 
force between D1 
and D5 and between F and NS5 in this bound state, we do have an attractive long-range force acting between F and D5 or between D1 and NS5 
(for example, see \cite{Ouyang:2014bha}). We,  therefore, expect a net attractive force between (F, D1) and (NS5, D5). This then appears to 
imply the existence of only the 1/2 BPS non-threshold ((F, D1), (NS5, D5)) bound state and not the 1/4 BPS threshold one, when (F, D1) is placed 
along one of the five world-volume spatial directions of (NS5, D5). 

It would be very surprising if this turns out to be true because it would imply a potential problem with the conjectured quantum SL(2, Z) symmetry 
of Type IIB string theory given the existence of (D1, D5) and (F, NS5) quantum mechanically\footnote{The low energy classical U-duality SL(2, R) 
symmetry can always be used to generate the classical solution 1/4 BPS ((F, D1), (NS5, D5)) from the given classical 1/4 BPS (D1, D5) or (F, NS5).}.   
In this paper, we will resolve this puzzle and find that there indeed exists a new type of ((F, D1), (NS5, D5)) bound state preserving 1/4 of the 
spacetime supersymmetry and both (D1, D5) and (F, NS5) states can be obtained from it as special cases.  Unlike the 1/2 BPS ((F, D1), (NS5, D5)) and 
also as anticipated, 
we expect certain constraint on the vacua, which manifest themselves in terms of the VEVs of the dilaton $\phi$ and RR 0-form potential $\chi$, i.e., 
$\phi_0$ and $\chi_0$.  It turns out that when $\phi_0$ and $\chi_0$ satisfy certain condition, the 1/4 BPS threshold ((F, D1), (NS5, D5)) bound state 
indeed exists. We also demonstrate that under precisely these conditions, there is no-force acting between (F, D1) and (NS5, D5) in this threshold 
bound state, lending support to the existence of this state. 

The main purpose of the present paper is to address the aforementioned puzzle and to construct such a state in type IIB string theory. To construct 
this state we need to begin with a known threshold 1/4 BPS (D1, D5) state or (F, NS5) state and apply SL(2, R) transformation and the charge quantizations
of the various constituent branes.  We will also spell out some peculiarities in the construction not encountered before in similar constructions. Also for
completeness we will also construct another 1/4 BPS bound state of the same form but here (F, D1) strings
will not be along (NS5, D5) brane spatial directions, but will be perpendicular to them. In this case, we will
start with the known 1/4 BPS (F, D5) bound state solution of type IIB string theory and apply an SL(2,R) rotation
and charge quantization to it. The charge relation in this case will be very similar to that of the 1/2 BPS ((F, D1), (NS5, D5)) solution
we constructed before\cite{Lu:1999uc}.  This will also serve as a contrast between these two kinds of 1/4 BPS states. 

This paper is organized as follows. In the next section we briefly discuss SL(2,R) symmetry of low energy type IIB
string theory and fix our notations and conventions. In section 3, we construct SL(2,Z) multiplet of 1/4 BPS ((F, D1), (NS5, D5))
solution starting from (D1, D5) solution using SL(2, R) and then imposing charge quantizations of constituent branes. We also derive 
the condition for the existence of such a state. 
In section 4, we consider a probe (F, D1)-string placed in (NS5, D5) brane background and derive the no-force condition. It turns out that 
the no-force condition is precisely the one 
derived in section 3. In section 5, we construct the other 1/4 BPS ((F, D1), (NS5, D5)) solution
starting from 1/4 BPS (F, D5) solution. Our conclusion is given in section 5.

\section{Type IIB supergravity and SL(2,R) symmetry} 
In this section we briefly review the low energy effective Lagrangian of type IIB string theory (type IIB supergravity
\cite{Schwarz:1983qr})
and mention how it can be written in an SL(2, R) invariant form. This will also fix our notations and conventions.     
The bosonic part of the Lagrangian has the form (see for example, \cite{Lu:1999bw}),
 \bea\label{lag}
 {\cal L}_{\rm IIB} &=& R \ast \mathbf{1} - \frac{1}{2} \ast d \phi \wedge d\phi - \frac{1}{2} e^{2 \phi} \ast d\chi \wedge d\chi - 
\frac{1}{4} \ast H_5\wedge H_5 \nn
&\,& - \frac{1}{2} e^{- \phi} \ast F^{\rm NS}_3 \wedge F^{\rm NS}_3 - \frac{1}{2} e^\phi \ast \tilde F_3\wedge \tilde F_3 
- \frac{1}{2} B_4 \wedge F^{RR}_3 \wedge F^{\rm NS}_3,
\eea 
where $\ast \mathbf {1}$ stands for the 10 dimensional volume-form, $R$, the Ricci scalar, $\phi$, the dilaton, 
$F^{\rm NS}_3$,  the NSNS 3-form field strength, $\chi$, the RR 0-form potential, $F^{\rm RR}_3$, the RR 3-form field strength 
and $B_4$, the RR 4-form potential. In the above $\ast$ denotes the spacetime Hodge-dual, and
\bea\label{fieldd}
&&F^{\rm NS}_3 = d A^{\rm NS}_2, \quad \tilde F_3 = d A^{\rm RR}_2 + \chi F^{\rm NS}_3,\nn
&& H_5 = d B_4 - \frac{1}{2} A^{\rm RR}_2 \wedge F^{\rm NS}_3 + \frac{1}{2} A^{\rm NS}_2 \wedge F^{\rm RR}_3.\eea
In eq.\eqref{lag}, we have included $H_5$ in the Lagrangian for the purpose of equations of motion for the other 
fields and at the end its self-dual relation $\ast H_5 = H_5$ is imposed by hand.

From the above Lagrangian, we have the following equations of motion
\bea\label{eom}
R_{MN} &=& \frac{1}{2} \partial_M \phi \partial_N \phi + \frac{1}{2} e^{2\phi} \partial_M \chi \partial_N \chi + \frac{1}{96} H^2_{ MN}\nn
               &\,&      + \frac{1}{4} e^\phi \left[\left(\tilde F_3\right)^2_{MN} - \frac{1}{12} g_{MN} \left(\tilde F_3\right)^2  \right] \nn
                 &\,&     + \frac{1}{4} e^{-\phi} \left[\left( F^{\rm NS}_3\right)^2_{MN} - \frac{1}{12} g_{MN} \left( F^{\rm NS}_3\right)^2 \right] ,\nn                
d \ast d \phi &=& - e^{2\phi} \ast d\chi \wedge d\chi - \frac{1}{2} e^\phi \ast \tilde F_3 \wedge \tilde F_3 + \frac{1}{2} e^{-\phi} 
\ast F^{\rm NS}_3\wedge F^{\rm NS}_3, \nn
d\left(e^{2\phi} \ast d\chi\right) &=&   - e^\phi \ast\tilde F_3 \wedge F^{\rm NS}_3,\quad d\ast H_5 = - F^{\rm RR}_3\wedge F^{\rm NS}_3\, \, \,({\rm with}\, \ast H_5 = H_5), \nn
d\left(e^\phi \ast \tilde F_3\right) &=& H_5 \wedge F^{\rm NS}_3,\quad d\left(e^{-\phi} \ast F^{\rm NS}_3 + \chi e^\phi \ast\tilde F_3\right) = - H_5 \wedge F^{\rm RR}_3.       
\eea
In order to see the manifest SL(2,R) symmetry, we now try to re-express the Lagrangian \eqref{lag} and the fields \eqref{fieldd} in an 
SL(2,R) invariant or covariant form. Note that the Einstein frame metric and the RR 5-form are invariant under SL(2,R). The totally 
antisymmetric tensor $\epsilon^{ij}$, where $i, j = 1, 2$ and $\epsilon^{12} = 1$, is SL(2,R) invariant. As usual, we define $F_3^{(1)} = F^{\rm NS}_3$ 
and $F^{(2)}_3 = F^{\rm RR}_3$.  We also define a 3-form vector $F_3$ and a $2\times 2$ scalar matrix ${\cal M}$ as
\be \label{sl2r}
F_3 = \left (\begin{array}{c} 
F^{(1)}_3\\
F^{(2)}_3\end{array}\right) ,\qquad   {\cal M} = \left(\begin{array}{cc} 
                                                                           \chi^2 + e^{- 2\phi} & \chi\\
                                                                           \chi & 1 \end{array}\right) e^\phi.
                                                                           \ee
 With the above, the Lagrangian can be re-expressed as
 \bea \label{sl2ra}
 {\cal L}_{\rm IIB} &=& R \ast \mathbf{1} - \frac{1}{4} \ast H_5 \wedge H_5 + \frac{1}{4} Tr \ast d {\cal M} \wedge d{\cal M}^{-1} - 
\frac{1}{2} \ast F_3^T \wedge {\cal M} F_3\nn
 &\,& + \frac{1}{4} \epsilon^{ij} B_4\wedge F^{(i)}_3 \wedge F^{(j)}_3.
 \eea
 This is manifestly SL(2,R) invariant if we have
 \be\label{transf}
 {\cal M} \to \Lambda{\cal M} \Lambda^T, \quad F_3 \to \left(\Lambda^{-1}\right)^T F_3, \quad  g_{MN} \to g_{MN},
 \ee
 where
 \be \label{epsilon}
 \Lambda^{il}\Lambda^{jk} \epsilon^{lk} = \epsilon^{ij},
 \ee
 defines $\Lambda$ as a $2\times 2$ SL(2,R) matrix.  In general, we can have
 \be
 \Lambda = \left (\begin{array}{cc}
                              \a & \b\\
                              \gamma & \d \end{array} \right),\quad {\rm with} \quad \a\d - \b\gamma = 1.
                              \ee
If we express the dilaton $\phi$ and the RR 0-form potential $\chi$ in terms of $\lambda = \chi + i e^{-\phi}$, 
it will transform fractional linearly under SL(2,R) as
\be\label{lambdat}
\lambda \to \lambda' = \frac{\a \lambda + \b}{\gamma\lambda + \d}.
\ee
The $H_5$ in \eqref{fieldd} can now be written as,
\be \label{h5}
H_5 = d B_4 + \frac{1}{2} \epsilon^{ij} A_2^{(i)} \wedge F^{(j)}_3, \quad d \ast H_5 = \frac{1}{2} \epsilon^{ij} F_3^{(i)}\wedge F_3^{(j)},
\ee
which is also SL(2,R) invariant as expected.       

\section{1/4 BPS ((F, D1), (NS5, D5)) from (D1, D5)}

In this section we will construct an SL(2,Z) multiplet of states in the form of 1/4 BPS threshold bound states ((F, D1), (NS5, D5)) starting from the
well-known 1/4 BPS threshold classical solution (D1, D5) of type IIB string theory by an SL(2,R) transformation and then imposing the 
charge quantizations of the different constituent branes. The field configuration for the (D1, D5) state (for example, see \cite{Lu:2009tw}) is, 
\bea\label{d1d5}
ds^2 &=& H_1^{-3/4} H_5^{-1/4} \left(- dt^2 + (d x^5)^2\right) + H_1^{1/4} H_5^{- 1/4} \sum^4_{i = 1} (d x^i)^2 \nn
          &\,& + H_1^{1/4} H_5^{3/4} \left( dr^2 + r^2 d\Omega_3^2\right),\nn
    e^{2\phi} &=& H_1/H_5,\nn
    F^{(2)}_3 &=& \frac{2 Q_1}{H_1^2 r^3} dt \wedge d x^5 \wedge d r + 2 Q_5 \epsilon_3,
    \eea
    where the two harmonic functions $H_{1/5} = 1 +  Q_{1/5}/r^2$, $\epsilon_3$ is the volume 3-form of a unit 3-sphere and 
parameters $Q_1, Q_5$ are related to the D1 and D5 charges 
    as
    \be 
    e_1 = \frac{2 Q_1 \Omega_3 V_4}{\sqrt{2} \kappa_0}, \qquad g_5 = \frac{2 Q_5\Omega_3}{\sqrt{2} \kappa_0},
    \ee
    with $\Omega_3 = 2\pi^2$ the volume of unit 3-sphere, $\sqrt{2} \kappa_0 = (2\pi)^{7/2} \alpha'^2$ and $V_4 = \int dx^1 \wedge dx^2 \wedge dx^3 \wedge d x^4$. 
The D-strings are along $x^5$ direction but delocalized along $1, 2, 3 , 4$-directions while the D5-branes are along $x^1, x^2, \cdots, x^5$ directions.  
In the above, we have set the asymptotic value $\phi_0 = 0, \chi_0 = 0$.   
    
The equation of motion for $A_2^{(i)}$ can now be obtained from the SL(2,R) invariant Lagrangian \eqref{sl2ra} as
\be\label{f3}
d ({\cal M}^{ij} \ast F_3^{(j)}) = -  \epsilon^{ij} H_5 \wedge F^{(j)}_3,
\ee
where we have used (\ref{h5}) and $\ast H_5 = H_5$.  This one agrees with the two equations in the last line of 
\eqref{eom}.  This equation can also be re-written as
\be\label{current}
d \left({\cal M}^{ij} \ast F^{(j)}_3 + \epsilon^{ij} B_4\wedge F^{(j)}_3 - \frac{1}{2} A^{(i)}_2\wedge A_2^{(j)} \wedge F^{(j)}_3\right) = 0,
\ee
from which we can define the electric-like (F, D1) string charge $e_1^T = (e_1^{(1)}, e_1^{(2)})$ as 
\be\label{e-charge}
e_1^{(i)} = \frac{1}{\sqrt{2} \kappa_0} \int_{R^4 \times S^3_\infty} {\cal M}^{ij} \ast F^{(j)}_3 + \epsilon^{ij} B_4\wedge F^{(j)}_3 - 
\frac{1}{2} A_2^{(i)}\wedge A^{(j)}_2\wedge F^{(j)}_3,
\ee
while the magnetic-like (NS5, D5) brane charge $g_5^T = (g_5^{(1)}, g_5^{(2)})$ is defined as
\be\label{m-charge}
g_5^{(i)} = \frac{1}{\sqrt{2} \kappa_0} \int_{S^3_\infty} F_3^{(i)}.
\ee 
For the present case, the Chern-Simons terms in the electric charge never contribute and so we drop them from now on.  
Given the SL(2,R) transformations of ${\cal M}$ and $F$ in
 (\ref{transf}), we have the electric-like charge $e_1$ and magnetic-like charge $g_5$ to transform as
 \be\label{charge-transf}
 e_1 \to \Lambda e_1, \qquad g_5 \to (\Lambda^{-1})^T g_5.
 \ee   
The standard choice for $\Lambda$ is 
\be\label{Lambda}
\Lambda = \left(\begin{array}{cc}
e^{ - \phi_0} \cos\alpha + \chi_0 \sin\alpha & - e^{-\phi_0} \sin\alpha + \chi_0 \cos\alpha\\
\sin\alpha & \cos\alpha \end{array}\right) e^{\phi_0/2}.
\ee 
Since (D1, D5) is our initial classical configuration with classical D1 charge $\Delta^{1/2}_1$ and classical D5 charge $\Delta^{1/2}_5$, respectively,  we have 
\be\label{istate}
e_1 = \left(\begin{array}{c}
0 \\
\Delta_1^{1/2} \end{array}\right) e_{10},  \qquad g_5 = \left(\begin{array}{c}
0 \\
\Delta_5^{1/2} \end{array}\right) g_{50},
\ee
where $e_{10}$ and $g_{50}$ are the unit charges for strings and 5-branes, respectively. After the SL(2,R) transformation, 
we should have the transformed classical charges and after imposing their respective quantizations, we end up with 
$\bar e^T_1 = (p, q) e_{10}$,  and  $\bar g_5^T = (m, n) g_{50}$ with $(p, q)$ and $(m, n)$ each being a pair of integers. Using
these quantization conditions we get, 
\bea\label{cossinalpha}
& & \cos\alpha = q e^{-\phi_0/2} \Delta_1^{-1/2} = (m\chi_0 + n) e^{\phi_0/2} \Delta_5^{-1/2},\nn
& & \sin\alpha = (q\chi_0 - p)e^{\phi_0/2} \Delta_1^{-1/2} = -m e^{-\phi_0/2} \Delta_5^{-1/2} .
\eea
Using \eqref{cossinalpha} we get,
\bea\label{deltatanalpha} 
& & \Delta_1 = q^2 e^{-\phi_0} + (q \chi_0 - p)^2 e^{\phi_0},  \qquad \Delta_5 = m^2 e^{-\phi_0} + (m \chi_0+n)^2 e^{\phi_0},\nn
& & \tan\alpha = \frac{(q\chi_0 - p)}{q} e^{\phi_0} = - \frac{m}{(m\chi_0 + n)} e^{-\phi_0},
\eea
and from the last relation we have,
\be\label{modulicondition}
(m\chi_0+n)(q\chi_0-p) = -mq \, e^{-2\phi_0},
\ee
which can actually be expressed as an SL(2, Z) invariant form\footnote{We thank one of our referees for pointing this out to us.}
\be\label{sl2zif}
(p, q) \epsilon^T {\cal M}_0 \left(\begin{array}{c}
m \\
n \end{array}\right) = 0,
\ee
where $\epsilon$ denotes the SL (2, Z) invariant antisymmetric matrix\footnote{Note also $\Lambda^T \epsilon^T \Lambda = \epsilon^T$ in addition to $\Lambda \epsilon \Lambda^T = \epsilon$ defined in \eqref{epsilon}.}  defined in \eqref{epsilon}.  
Eq.\eqref{modulicondition} can be satisfied with arbitrary $\phi_0$ and $\chi_0$, only if $qm=0$, $pn=0$ and also $(qn-pm)=0$. However,
we note that these relations are inconsistent even for the (D1, D5) configuration\footnote{
\label{fntwo} Even though we have chosen the initial
(D1, D5) configuration \eqref{d1d5} with $\phi_0=\chi_0=0$, it can have arbitrary $\phi_0$ and constant $\chi_0$ as can be seen from 
the corresponding 
equations of motion \eqref{eom}, even after imposing the charge quantizations of D1 and D5, respectively.} for which $q$ and $n$ are both 
non-zero. Thus we conclude that for SL(2,Z) multiplet of 1/4 BPS 
((F, D1), (NS5, D5)) bound state to exist, $\phi_0$ and/or $\chi_0$ can not be arbitrary but must take specific values such that
they satisfy the condition \eqref{modulicondition}. We would like to emphasize that in this bound state $(p,q)$ strings (F, D1) form a 
threshold
bound state with $(m,n)$ 5-branes (NS5, D5) which is implied from the original (D1, D5) metric since the total energy is simply the sum
of those of the delocalized (F, D1) and (NS5, D5). Therefore, there is no interaction between (F, D1) and (NS5, D5) just as there is
no interaction between the original D1 and D5. This is far from obvious since in general we do expect interaction between (F, D1) and
(NS5, D5) since we have attractive interaction between F and D5 and also between D1 and NS5\cite{Ouyang:2014bha}. We will demonstrate 
explicitly in the following section by computing
the force experienced by the $(p,q)$ string in the $(m,n)$ 5-brane background that indeed the no-force condition is precisely the 
same as the condition \eqref{modulicondition}. 

One important point to note about \eqref{modulicondition} is that, the quantized charges of various objects in this bound state appear
to get fixed by the moduli $\phi_0$ and $\chi_0$. But that is not desirable and not right since the values of the moduli are
given and the quantization condition is imposed later and therefore charges should be able to take any quantized value
independent of $\phi_0$ and $\chi_0$. In order to
achieve this we re-write $(p,q) = k(a,b)$ and $(m,n)=k'(a',b')$, where $(a,b)$, $(a',b')$ are two pairs of co-prime integers and
$k$, $k'$ are integers but not necessarily co-prime. This is because (F, D1) and (NS5, D5) form only threshold bound state just like
original (D1, D5).  For the non-degenerate configuration $kk'\neq 0$ and so, substituting these values
of $(p,q)$ and $(m,n)$ in \eqref{modulicondition}, we find that $\phi_0$ and $\chi_0$ are not linked with $(p,q)$ and $(m,n)$, but rather
with $(a,b)$ and $(a',b')$. Actually the charges $(p,q)$ and $(m,n)$ can take almost any arbitrary integer values except for the 
degenerate $k k'  = 0$ cases.

For the concerned state to remain 1/4 BPS,  we must have $k k' \neq 0$.  The degenerate $k \neq 0, k' = 0$ or $k = 0, k' \neq 0$ case implies 
that the underlying state is either delocalized 1/2 BPS  $(p, q)$-strings or 1/2 BPS $(m, n)$ 5-branes. However, for either of these two cases, 
both (\ref{cossinalpha}) and (\ref{deltatanalpha}) are not well-defined and for this reason we don't expect (\ref{modulicondition}) to hold 
good. This is entirely consistent since (\ref{modulicondition}) is the constraint for 1/4 BPS state, not for 1/2 BPS state. The condition 
$k k' \neq 0$ implies that none of  the two integers in either $(p, q)$ or $(m, n)$ pair can vanish.  We remark here that 
(\ref{modulicondition}) rules out the existence of (F, D5) or (D1, NS5). For the former, we have $p, n \neq 0, q = m = 0$ and (\ref{modulicondition}) 
gives $pn = 0$, contradicting our assumption, while for the latter, we have $q, m \neq 0, p = n =0$ and (\ref{modulicondition}) gives 
$e^{- 2\phi_0} = - \chi_0^2$, impossible to hold.  This is also consistent since either of these two is a 1/2 BPS state. To make our following 
discussion definite, we are assuming $k k' \neq 0$ from now on.  Then  (\ref{modulicondition}) 
becomes, in terms of $(a, b)$ and $(a', b')$, as
\be\label{keymodulicondition}
(a'\chi_0+b')(b\chi_0- a) = - a' b \, e^{-2\phi_0}. 
\ee
We also have now
\bea\label{newcossin}
& & \cos\alpha = b \, e^{-\phi_0/2}\tilde\Delta_1^{-1/2} = (a'\chi_0 + b') e^{\phi_0/2} \tilde\Delta_5^{-1/2},\nn
& & \sin\alpha = (b\chi_0 - a)e^{\phi_0/2} \tilde\Delta_1^{-1/2} = -a' e^{-\phi_0/2} \tilde\Delta_5^{-1/2},
\eea
where 
\be\label{newdelta} 
 \tilde\Delta_1 = b^2 e^{-\phi_0} + (b \chi_0 - a)^2 e^{\phi_0},  \qquad \tilde\Delta_5 = a'^2 e^{-\phi_0} + (a' \chi_0+b')^2 e^{\phi_0}.
\ee

We would like to address first a few subtleties regarding the generation of (D1, D5) (or (F, NS5)) solution with non-zero 
moduli ($\chi_0,\,\phi_0 \neq 0$) from the same solution with zero moduli ($\chi_0=\phi_0=0$) before we proceed to discuss the 
general 1/4 BPS state including certain special cases using (\ref{keymodulicondition}).  We note that the SL(2,R) matrix given 
in \eqref{Lambda} can be decomposed as,
\be\label{Lambdadecom}
\Lambda = \left(\begin{array}{cc} 
                     e^{- \phi_0} & \chi_0\\
                     0 & 1 \end{array}\right) e^{\phi_0/2}\,\,\, \left(\begin{array}{cc}
\cos\alpha & -\sin\alpha\\
\sin\alpha & \cos\alpha \end{array}\right) \equiv \Lambda_0 \Lambda_R, 
\ee   
where $\Lambda_R$ is the SO(2) rotation subgroup of SL(2,R). $\Lambda_R$ actually generates new solutions, i.e., starting from (D1, D5) with 
$\phi_0 = \chi_0 = 0$ it generates another solution with again zero moduli since $\Lambda_R {\bf I} \Lambda_R^T = {\bf I}$.
On the other hand, $\Lambda_0$ generates (D1, D5) solution with non-zero moduli ($\chi_0,\,\phi_0 \neq 0$) from the same solution with zero moduli ($\chi_0=\phi_0=0$).
The last statement can be justified by looking at the relation ${\cal M}_0 = \Lambda_0 {\bf I} \Lambda_0^T$, where ${\cal M}_0$ is the
matrix given in \eqref{sl2r} with the asymptotic values of the moduli and ${\bf I}$ is the 2 $\times$ 2 identity matrix which is
nothing but ${\cal M}_0$ with the moduli put to zero.  Since there are no F-strings and/or NS5-branes generated under the special 
SL(2,R) transformation of $\Lambda  = \Lambda_0$, we should not blindly enforce the charge quantization for either of these two kinds of 
branes. In other words, we don't have  (\ref{cossinalpha}) and (\ref{deltatanalpha})  for this case and so (\ref{modulicondition}) also does 
not hold.  So $\phi_0$ and $\chi_0$ can be arbitrary for (D1, D5), not constrained by (\ref{modulicondition}). This same discussion 
holds true if we replace (D1, D5) by (F, NS5) in the above. In general, (D1, D5) (or (F, NS5)) are physically different with different 
$\phi_0, \chi_0$ except for the special case $\phi_0 = 0, \chi_0 = {\rm interger}$. This latter special case is physically equivalent to 
(D1, D5) (or (F, NS5)) with $\phi_0 = \chi_0 = 0$ since these two are related by an SL(2,Z) transformation $ \Lambda_0$ with $\phi_0 = 0, \chi_0 = {\rm integer}$. 

We now come to give a general discussion on the non-degenerate 1/4 BPS threshold ((F, D1), (NS5, D5)) state including certain special cases 
using (\ref{keymodulicondition}) as well as (\ref{newcossin}). Since the pair $(a, b)$ (or $(a', b')$) are co-prime, so at most one of two co-prime integers 
can vanish.  We will discuss case by case in the following.

\noindent
{\bf Case 1 $a = 0, b\neq 0$:}. For this, from  (\ref{keymodulicondition}), we have $a' e^{- 2\phi_0} = - \chi_0 (a' \chi_0 + b')$.  We have three 
subcases to consider: 1a) $a' = 0, b' \neq 0$, then we must have $\chi_0 = 0$ with $\phi_0$ arbitrary.   This subcase is just 1/4 BPS (D1, D5) 
with $\chi_0 = 0$ and arbitrary $\phi_0$. For 1/4 BPS (D1, D5), our previous discussion says that it can exist with arbitrary $\chi_0$ and $\phi_0$.  
The restriction $\chi_0 = 0$ obtained here is due to the charge quantization condition imposed for F-string and NS5-brane when we consider 1/4 BPS 
((F, D1), (NS5, D5)). However, this condition is actually irrelevant when we merely consider (D1, D5) and should be dropped. 1b) $a' \neq 0, b' =0$. 
This subcase is impossible since it requires $e^{- 2\phi_0} = - \chi_0^2$. This just tells us that 1/4 BPS (D1, NS5) does not exist as we discussed previously. 
1c) $a' \neq 0, b' \neq 0$. Now we have $e^{- 2\phi_0} = - \chi_0 (\chi_0 + b'/a')$, which gives a restriction on $\chi_0$ as $- b'/a' < \chi_0 < 0$ if $a' b' > 0$ 
or $ 0 < \chi_0 < - b'/a'$ if $a'b' < 0$.  This subcase gives 1/4 BPS (D1, (NS5, D5)) if the moduli $\phi_0$ and $\chi_0$ satisfy these constraints.  

\noindent
{\bf Case 2 $a \neq 0,  b = 0$:} For this, we have $a' \chi_0 + b' = 0$ from  (\ref{keymodulicondition}).  So $a'$ cannot be zero and we have two 
subcases to consider. 2a) $a' \neq 0, b' =0$. This gives $\chi_0 = 0$.  We have now 1/4 BPS (F, NS5) with $\chi_0  = 0$ and arbitrary $\phi_0$. By 
the same argument discussed in subcase 1a) for (D1, D5), this case actually works for arbitrary $\chi_0$ and $\phi_0$. 2b) $a' \neq 0, b' \neq 0$. 
We then have $\chi_0 = - b'/a'$. This gives 1/4 BPS (F, (NS5, D5)) with $\chi_0 = - b'/a'$ and arbitrary $\phi_0$.  

\noindent
{\bf Case 3 $a \neq 0, b \neq 0$:}  We have three subcases to consider. 3a) For $a' = 0, b' \neq 0$, we have $\chi_0 = a/b$ with $\phi_0$ arbitrary.  
So we have 1/4 BPS threshold ((F, D1), D5) with $\chi_0 = a/b$ and arbitrary $\phi_0$. 3b) $a' \neq 0, b' = 0$. For this subcase, we have 
$e^{- 2\phi_0} = - \chi_0 (\chi_0 - a/b)$ with the constraint on $\chi_0$ as $ 0 < \chi_0 < a/b$ if $a b > 0$ or $ a/b < \chi_0 < 0$ if $ a b < 0$.  
We now have 1/4 BPS ((F, D1), NS5) with the moduli $\phi_0$ and $\chi_0$ satisfying the conditions discussed. 3c) $a' \neq 0$ and $b' \neq 0$. 
This is the generic  non-degenerate 1/4 BPS ((F, D1), (NS5, D5)) case with its moduli $\chi_0$ and $\phi_0$ satisfying 
\be\label{keyonemodulicondition}
e^{-2\phi_0} = - \left(\chi_0+\frac{b'}{a'}\right)\left(\chi_0- \frac{a}{b}\right) . 
\ee 
For this subcase, there is an interesting and very special case with $\phi_0 = \chi_0 = 0$ for which we have $b' a/a' b = 1$ from (\ref{keyonemodulicondition}).  
Since both $(a', b')$ and $(a, b)$ are co-prime, this must imply $(a', b') = (a, b)$.

In all the above discussion,  there are two additional constraints implied by (\ref{newcossin}) for $\chi_0$, $(a, b)$ and $(a', b')$. They are
\be\label{additionalconstraint}
a' (b \chi_0 - a) \le 0, \quad b (a' \chi_0 + b') > 0 \quad {\rm or} \quad a' (b \chi_0 - a ) < 0, \quad b(a' \chi_0 + b') \ge 0.
\ee

For the purpose of comparison, we remark here that the 1/2 BPS ((F, D1), (NS5, D5)) puts less constraint on the moduli $\phi_0, \chi_0$ but more 
on the charges $(p, q) = k (a, b), (m, n) = k' (-b, a)$ with both $(k, k')$ and $(a, b)$ co-prime. But for the 1/4 BPS ((F, D1), (NS5, D5)), on the contrary, 
we have more constraint on the moduli satisfying (\ref{keyonemodulicondition}) but less on the charges $(p, q) = k (a, b), (m, n) = k' (a', b')$ with  
$k k' \neq 0$ and  $(a, b)$ and $(a', b')$  being two pairs of co-prime integers.  

We  give below the most general 1/4 BPS SL(2,Z) invariant ((F, D1), (NS5, D5)) bound state,            
\bea\label{fd1ns5d5}
ds^2 &=& {\cal H}_1^{-3/4} {\cal H}_5^{-1/4} \left(- dt^2 + (d x^5)^2\right) + {\cal H}_1^{1/4} {\cal H}_5^{- 1/4} \sum^4_{i = 1} (d x^i)^2 \nn
          &\,& + {\cal H}_1^{1/4} {\cal H}_5^{3/4} \left( dr^2 + r^2 d\Omega_3^2\right),\nn
    e^{2\phi} &=& \frac{\left( \delta^2{\cal H}_1 + \gamma^2{\cal H}_5 \right)^2}{{\cal H}_1{\cal H}_5}, \qquad
    \chi = \frac{\beta \delta{\cal H}_1 + \alpha \gamma{\cal H}_5}{\delta^2 {\cal H}_1 + \gamma^2 {\cal H}_5},\nn 
    F^{(1)}_3 &=& - \gamma\left[\frac{2  Q_1}{{\cal H}_1^2 r^3} dt \wedge d x^5 \wedge d r + 2 Q_5 \epsilon_3\right],\nn 
    F^{(2)}_3 &=& \alpha \left[\frac{2  Q_1}{{\cal H}_1^2 r^3} dt \wedge d x^5 \wedge d r + 2 Q_5 \epsilon_3\right].
    \eea
Here ${\cal H}_{1/5} = 1 + Q_{1/5}/r^2$ are the new harmonic functions which have the same form as the old harmonic functions but now with
 $ Q_{1/5} = \Delta^{1/2}_{1/5} Q_0$, where $\Delta_{1/5}$ are as given in \eqref{deltatanalpha} and $Q_0 = \sqrt{2}\kappa_0 Q_0^5 /(2 \Omega_3) = \alpha'$ with 
$Q_0^p = (2\pi)^{(7 - 2 p)/2}\alpha'^{(3 -p)/2}$ \cite{Lu:1999uc}. Also $\alpha,\,\beta,\,\gamma,\,\delta$ are the
various components of the SL(2,R) matrix \eqref{Lambda} and can be read off from \eqref{cossinalpha} as,
\bea\label{abcd}
& & \alpha = \left[qe^{-\phi_0} + \chi_0(q\chi_0 -p)e^{\phi_0}\right]\Delta_1^{-\half} = n \Delta_5^{-\half},\nn  
& & \beta = p \Delta_1^{-\half} = \left[me^{-\phi_0} + \chi_0(m\chi_0+n)e^{\phi_0}\right]\Delta_5^{-\half},\nn
& & \gamma = (q\chi_0 - p) e^{\phi_0} \Delta_1^{-\half} = -m\Delta_5^{-\half},\nn
& & \delta = q \Delta_1^{-\half} = (m\chi_0+n)e^{\phi_0}\Delta_5^{-\half}.
\eea
We note that since the Einstein frame metric is invariant under SL(2,R) transformation, the SL(2,R) transformed metric in \eqref{fd1ns5d5}
has the same form as the original (D1, D5) metric given in \eqref{d1d5} except that the quantity $Q_{1/5}$ related to the charge has been
changed due to proper charge quantization. The original solution has $\chi=0$, but after SL(2,R) transformation a $\chi$ has been generated.
The dilaton has also been transformed. We can easily check that asymptotically as $r \to \infty$, ${\cal H}_{1/5} \to 1$ and therefore 
$\phi \to \phi_0$ and $\chi \to \chi_0$ as expected. We also note that there is no NSNS 3-form in the initial (D1, D5) solution, but
it has been generated by SL(2,R) transformation. However, no 5-form field has been generated as opposed to the 1/2 BPS ((F, D1), (NS5, D5))
state\cite{Lu:1999uc}.

Following the steps given in \cite{Lu:1999uca} and \cite{Lu:1993vt}, the string-frame tension of the bound state ((F, D1), (NS5, D5)) can be computed from 
the ADM mass per unit 5-brane volume as
\be\label{tension}
 T_5 (k, k'; a, b; a', b') = \frac{T^5_{0}}{g} \left (k \sqrt{b^2 + (b\chi_0 - a)^2 g^2}  + k' \sqrt{a'^2 g^{-2} + (a' \chi_0 + b')^2}\right), 
 \ee
where $T_0^p = 1/[(2\pi)^p \alpha'^{(p + 1)/2}]$ is the $p$-brane tension unit  and $g = e^{\phi_0}$ is the string coupling.  Here we have also set 
$(p, q) = k (a, b), (m, n) = k' (a', b')$
with $(a, b)$ and $(a', b')$ both co-prime.  This tension is the sum of contributions from $(p, q)$ strings and $(m, n)$ 5-branes, respectively, 
and this is actually determined by the form of the metric given in (\ref{fd1ns5d5}), which is the same in form as that of (D1, D5), mentioned earlier.  
This tension gives all expected properties, for example, the dependence of string coupling for each kind of constituent branes. We will not discuss 
them explicitly here. 

\begin{table}
\begin{center}
 \begin{tabular}{|c|c|c|}   
\hline
 & Parallel & Transverse\\
 \hline
 Dp & D(p - 1) & D(p + 1)\\
 \hline
 F & W & F\\
 \hline
 W & F & W\\
 \hline
NS5 & NS5 & KK\\
\hline
 KK & KK & NS5\\
 \hline 
\end{tabular}
\caption{The T-duality rule along parallel or transverse direction to the branes or waves in Type II Theories.}\label{t1}
\end{center}
 \end{table}
 One can also get various 1/4 BPS descendant states from this one using T-dualities along the 5-brane worldvolume 
isometric directions or directions transverse to the 5-branes following the standard rule as given in Table \ref{t1}. In this table, KK and W stand for KK-monopole and waves, respectively.  So, for example, if we take a T-duality transformation along the 
string direction, we will end up with a 1/4 BPS ((W, D0), (NS5, D4))  and if we take T-duality along the 5-brane but not the string direction, 
we end up with a 1/4 BPS ((F, D2), (NS5, D4)) and so on. We can also take T-duality along one of transverse directions to the 5-branes for which we 
first need have one isometry along this T-dual direction. We then end up with 1/4 BPS ((F, D2), (KK, D6)).

\section{No-force condition}

As discussed in the Introduction, we, in general, expect an attractive interaction between the non-threshold $(p, q)$ strings and the non-threshold $(m,n)$ 5-branes 
due to the attractive force between F and D5 as well as between D1 and NS5.  If this interaction cannot be turned off, the 1/4 BPS threshold ((F, D1), (NS5, D5)) state 
cannot exist.  However, in the previous section, using the classical SL(2,R) transformation on a 1/4 BPS (D1, D5) state and then imposing the charge quantization 
for each kind of constituent branes, we seem to have obtained such 1/4 BPS threshold state when the moduli $\phi_0$ and $\chi_0$ satisfy the condition (\ref{modulicondition}) or
(\ref{keymodulicondition}) or (\ref{keyonemodulicondition}). This, therefore, suggests that the condition on the moduli just mentioned, actually  
plays a key role to change the nature of force, contrary to our naive understanding, such that the net force acting between (F, D1) and (NS5, D5) vanishes.  
Otherwise, what we have found in the previous section is simply wrong.  Therefore, to cross-check our finding on the 1/4 BPS threshold ((F, D1), (NS5, D5)) state,  
we need to explicitly compute the net force to see whether it can vanish and under what condition.  We find nicely that the condition 
for which the net force vanishes is precisely the one given in the previous section for the moduli $\phi_0$ and $\chi_0$. 

For this purpose,  let us place a $(p, q)$ string in an $(m, n)$ 5-brane background and see, under what condition the force between the two vanishes. 
The $(m, n)$ 5-brane background \cite{Lu:1998vh} is 
 \bea\label{mn5branebg}
 ds^2 &=& A_{(m, n)}^{1/4} \left(- dt^2 + d x^i d x^j \delta_{ij} \right) + A^{-3/4}_{(m, n)} \left(d r^2 + r^2 d\Omega^2_3\right),\nn
 e^{- \phi} & =& \frac{\Delta_5 \,A^{1/2}_{(m, n)} e^{\phi_0}}{m^2 e^{- \phi_0} + A_{(m, n)} e^{\phi_0} (m\chi_0 + n)^2},\nn
 \chi &=& \frac{\chi_0 \Delta_5\, A_{(m, n)} + mn e^{- \phi_0} (A_{(m, n)} - 1)}{m^2 e^{-\phi_0} + A_{(m, n)} e^{\phi_0} (m\chi_0 + n)^2},\nn
 F^{(1)}_3 &=& 2 m Q_0 \epsilon_3, \qquad F^{(2)}_3 = 2 n Q_0 \epsilon_3,
 \eea
 where the metric is in the Einstein frame and in the above
 \be\label{hf}
 A^{- 1}_{(m, n)} = 1 + \frac{Q_{(m, n)}}{ r^2}, \,\, {\rm with}\,\, Q_{(m, n)} = \Delta^{1/2}_5 Q_0,
 \ee  
 with $\Delta_5$ defined earlier for $(m, n)$ 5-brane in (\ref{deltatanalpha}) and $Q_0 = \sqrt{2}\kappa_0 Q_0^5 /(2 \Omega_3) $ 
(Note here $Q_0^p = (2\pi)^{(7 - 2 p)/2}\alpha'^{(3 -p)/2}$). 
 There are various forms of $(p, q)$ string action \cite{Townsend:1997kr, Cederwall:1997ts,Bergshoeff:2007ma} and for convenience 
we here use the bosonic part of $(p, q)$ string action which was first given in \cite{Bergshoeff:2006gs}.  Since $(p,q)$ string will not couple 
to $(m, n)$ 5-brane magnetic 3-form field strength, we only need to consider the part of the couplings of $(p, q)$ string  with the 
background metric, dilaton and the 0-form $\chi$ of the $(m, n)$ 5-brane.  It is given as
\bea\label{pqstringaction}
S &=& - T \int d^2 \sigma \sqrt{q^2 e^{-\phi} + (p - q\chi)^2 e^\phi} \, \sqrt{- h} \nn
&=& - T \int d^2\sigma {\bar \Delta_1 }^{1/2} \, \sqrt{- h} ,                                            
\eea
where 
\be\label{bardelta}
\bar \Delta_1 =  (p, q)^T {\cal M}^{-1} \left(\begin{array}{c} p\\q  \end{array}\right) = q^2 e^{-\phi} + (p - q\chi)^2 e^\phi.
\ee
Here  ${\cal M}$ is the scalar matrix defined in (\ref{sl2r}),  and $h = \det h_{\alpha\beta}$, with $h_{\alpha\beta}$, the induced world-sheet 
metric from spacetime Einstein metric $g_{MN}$ given in  (\ref{mn5branebg}) as
\be\label{inducedm}
h_{\alpha\beta} = \partial_\alpha X^M \partial_\beta X^N \, g_{MN},
\ee
where the worldsheet indices $\alpha, \beta = 0, 1$.  So from the second line of the action (\ref{pqstringaction}), it is clear that the action 
is manifestly SL(2,Z) invariant. 
The equation of motion for $X^M$ can be derived from the action (\ref{pqstringaction}) and after a lengthy calculation we obtain it as  
\be\label{eomnew}
 \frac{1}{\sqrt{- h}}\partial_\alpha \left(\sqrt{- h} h^{\alpha\beta} \partial_\beta X^M\right) +  V^M = 0,
 \ee 
where 
\bea\label{potential}
&&V^M =  \Gamma^{M}_{NP} h^{\alpha\beta} \partial_\alpha X^N \partial_\beta X^P \nn
&& \qquad\quad +  \frac{2 q  (q \chi - p) \partial_N \chi + ( (q\chi - p)^2 - q^2 e^{- 2\phi}) \partial_N \phi}{ 2 \, \bar \Delta_1\, 
e^{- \phi}}(h^{\alpha\beta} \partial_\alpha X^N \partial_\beta X^M - g^{NM}).
\eea
In the above, 
\be
\Gamma^M_{NP} = \frac{1}{2} g^{MQ} \left(\partial_N g_{QP} + \partial_P g_{Q N} - \partial_Q g_{NP}\right),
\ee
is the usual Christoffel symbol.  We now look for the conditions for which the force $V^M$  between $(p,q)$ string and  $(m, n)$ 5-brane 
vanishes when $(p, q)$ string is placed along one of the five spatial directions ($x^1, x^2, \cdots, x^5$, say $x^1$) of the 5-brane.  In other words, 
we are looking for the ``no-force'' condition. We also take the so-called static gauge for the 
string, namely,  $\tau = X^0, \sigma = X^1$ with string worldsheet coordinates $\sigma^\alpha = (\tau, \sigma)$. If string is parallel to 
5-brane and is static, vanishing force, 
i.e., $V^M = 0$, would then imply from (\ref{eomnew}) that it will remain static.  So, for now, all the $X^M$'s except $X^0 = \tau, X^1 =\sigma$ 
will be independent of $\tau, \sigma$.  We then have
\be\label{metric-string}
h_{\alpha\beta} =\partial_\alpha X^M \partial_\beta X^N g_{MN} = g_{11} \eta_{\alpha\beta},
\ee
where $g_{11} = A_{(m,n)}^{1/4}$ as given by the metric (\ref{mn5branebg}) and $\eta_{\alpha\beta} = (- 1, 1)$ the worldsheet flat metric.  
The relevant $\Gamma^M_{NP}$ for $V^M$ are $\Gamma^M_{00}$ and $\Gamma^M_{11}$  and their non-vanishing components  are
\be\label{affinec}
\Gamma^r_{00} = - \Gamma^r_{11} = \frac{1}{2} g^{rr} \partial_r g_{11}.
\ee
 Note that both $\chi$ and $\phi$ are only functions of $r$ and so automatically $V^M = 0$ for all $M$ except for $M = r$.
Now we need to check under what condition $V^r = 0$.  The first term on the right hand side of (\ref{potential}) can be expressed as
\be 
\Gamma^{M}_{NP} h^{\alpha\beta} \partial_\alpha X^N \partial_\beta X^P =  - g^{rr} g^{11} \partial_r g_{11} =  -  \frac{1}{4} g^{rr}  A_{(m, n)}^{- 1} \partial_r A_{(m, n)},
\ee
where in the last equality the metric component $g_{11}$ given in (\ref{mn5branebg}) has been used. From $\chi$ and $\phi$ given 
in (\ref{mn5branebg}), we also have
\bea\label{chi-phi}
\partial_r \chi &=& \frac{m (m\chi_0 + n)\,e^{- \phi_0}\,\Delta_5 \, \partial_r A_{(m, n)}}{\left[m^2 e^{- \phi_0} + A_{(m,n)} (m\chi_0 + n)^2 e^\phi_0\right]^2},\nn
\partial_r \phi &=& \frac{\left[A_{(m,n)} (m\chi_0 +n)^2 - m^2 e^{- 2\phi_0}\right] 
e^{\phi_0}}{2 \left[ m^2 e^{- \phi_0} + A_{(m,n)} (m\chi_0 + n)^2 e^\phi_0\right]} A^{-1}_{(m,n)} \partial_r A_{(m,n)}.
\eea 
We then have from \eqref{potential},
\be\label{forcer}
- V_r = \frac{1}{4} A^{-1}_{(m,n)} \partial_r A_{(m,n)} + \frac{1}{2}  \frac{2 q  (q \chi - p) \partial_r \chi + 
\left( (q\chi - p)^2 - q^2 e^{- 2\phi}\right) \partial_r \phi}{ 2 \left[(q\chi - p)^2 + q^2 e^{-2\phi}\right]},
\ee
where $V_r \equiv g_{rr} V^r$ and we have used $\bar\Delta_1$ given in (\ref{bardelta}).  At first look, the simplification of the right hand side of \eqref{forcer}
might appear difficult. But a careful examination reveals that a great simplification can be achieved if we first combine the first term on the right hand side of 
\eqref{forcer} with the term involving  $\partial_r \phi$ and then we add the term involving $\partial_r \chi$.  Let us see this in a bit detail. The first 
combination just described gives   
\bea\label{twots}
&&\frac{\partial_r A_{(m,n)} e^{\phi_0}}{2\bar \Delta_1 e^{ -\phi} \left[m^2 e^{-\phi_0} + A_{(m,n)} 
(m\chi_0 + n)^2 e^{\phi_0}\right]^3} \left\{q^2 m^2 \Delta_5^2 e^{- 4 \phi_0}\right.\nn
&&\left. + \left(m\chi_0 + n \right)^2  
 [(q \Delta_5 \chi_0 + q mn e^{- \phi_0} - p (m\chi_0 + n)^2 e^{\phi_0}) A_{(m,n)} - m (qn + pm) e^{- \phi_0}]^2  \right\},\nn
\eea   
where $\partial_r \phi$ given in (\ref{chi-phi}) has been used. 
The term involving $\partial_r \chi$ on the right hand side of (\ref{forcer}) can be expressed as 
\bea\label{chi}
&&\frac{q (q \chi - p) \partial_r \chi}{2 \bar \Delta_1 e^{-\phi}}=
 \frac{ q m(m\chi_0 + n)\Delta_5 e^{-\phi_0}\,\partial_r A_{(m,n)}}{ 2 \bar\Delta_1 e^{-\phi} [m^2 e^{-\phi_0} + A_{(m,n)} (m\chi_0 + n)^2 e^{\phi_0}]^3} \times\nn
&&\qquad\qquad\left[(q \Delta_5 \chi_0 + q mn e^{- \phi_0} - p (m\chi_0 + n)^2 e^{\phi_0}) A_{(m,n)} - m (qn + pm) e^{- \phi_0}\right].
\eea
Combining (\ref{twots}) with (\ref{chi}), we end up with
\be\label{finalsum}
V_r  = - \frac{\partial_r A_{(m, n)} e^{\phi_0} }{2 \bar\Delta_1 e^{- \phi} \left[m^2 e^{-\phi_0} + A_{(m,n)} (m\chi_0 + n)^2 e^{\phi_0}\right]^3}  \bar V_r,
\ee
where 
\bea\label{barV}
\bar V_r &=& \left\{ (m \chi_0 + n) \left[(q \Delta_5 \chi_0 + q mn e^{- \phi_0} - p (m\chi_0 + n)^2 e^{\phi_0}) A_{(m,n)} - m (qn + pm) e^{- \phi_0}\right]\right.\nn
&\,& \left. + qm \Delta_5 e^{-2 \phi_0}\right\}^2,
\eea
is a perfect square. 
So $V_r = 0$ amounts to setting $\bar V_r = 0$, which implies 
\bea\label{no-force}
&&(m\chi_0 + n) \left[q \Delta_5 \chi_0 + q mn e^{- \phi_0} - p (m\chi_0 + n)^2 e^{\phi_0}\right] = 0, \nn
&& qm \Delta_5 e^{- 2\phi_0} - m (m\chi_0 + n) (qn + mp) e^{-\phi_0} = 0.
\eea
The above two equations can be further simplified to give
\bea\label{no-forceone}
&&(m\chi_0 + n)^2 \left[qm e^{- 2\phi_0} + (m\chi_0 + n) (q\chi_0 - p)\right] = 0,\nn
&& m^2 \left[qm e^{- 2\phi_0} + (m\chi_0 + n) (q\chi_0 - p)\right]  = 0.
\eea
These two equations amount to requiring 
\be\label{keycondition}
qm e^{- 2\phi_0} + (m\chi_0 + n) (q\chi_0 - p) = 0,
\ee
which is nothing but the condition (\ref{modulicondition}) derived in the previous section for the existence of 1/4 BPS ((F, D1), (NS5, D5)) state.  
We can also set $(p, q) = k (a, b),\,\, (m, n) = k' (a', b')$ with again $(a, b)$ and $(a', b')$ being two pairs of co-prime integers. Here again the 
generic consideration of the force between (F, D1) and (NS5, D5) requires $k k' \neq 0$. Then the discussion of  the ``no-force'' condition will go 
exactly the same way as that for the existence of 1/4 BPS state given in the previous section and so we will not repeat it here. 
 
    So we conclude that when a $(p, q)$ string is placed in an $(m, n)$ 5-brane background with a given vacuum, i.e., with given $\phi_0$ and $\chi_0$, 
along one of the 5-brane world-volume spatial directions, one in general expects a net attractive force between them since we know that in general the 
forces between the constituent branes are either attractive or zero (the force between F and D5 as well as between D1 and NSNS5 is attractive while 
the others are zero) and therefore the end result is a non-threshold 1/2 BPS ((F, D1), (NS5, D5)) state as discussed in \cite{Lu:1999uc}. 
However, when the no-force condition (\ref{keycondition}) holds, then the end result is a 1/4 BPS threshold bound state ((F, D1), (NS5, D5)) 
as discussed in this paper.

\section{Another 1/4 BPS ((F, D1), (NS5, D5)) from 1/4 BPS (F, D5)}   

In this section, we will give the construction of another 1/4 BPS ((F, D1), (NS5, D5)) state starting from a known 1/4 BPS
(F, D5) state of type IIB string theory for completeness. Unlike in the previous case here $(p, q)$ string is perpendicular to
the $(m, n)$ 5-brane. However, the construction and the charge relation in this case are very similar to those of 1/2 BPS 
((F, D1), (NS5, D5)) state we constructed earlier in \cite{Lu:1999uc} and so we will be brief here. The 1/4 BPS (F, D$p$)
threshold bound states were constructed in \cite{Dey:2012tg}, so, we write down the (F, D5) solution from that reference, putting
$p=5$ in their eq.(2.6), in the following,
\bea\label{fd5}
ds^2 &=&  - H_1^{-3/4} H_5^{-1/4} dt^2 + H_1^{1/4} H_5^{- 1/4} \sum^5_{i = 1} (d x^i)^2 \nn
          &\,& + H_1^{-3/4} H_5^{3/4} (dx^6)^2 + H_1^{1/4} H_5^{3/4} \left( dr^2 + r^2 d\Omega_2^2\right),\nn
    e^{2\phi} &=& \frac{1}{H_1 H_5},\nn
    F^{(1)}_3 &=& \frac{Q_1}{H_1^2 r^2} dt \wedge d x^6 \wedge dr,\nn
    F^{(2)}_3 &=& Q_5 \epsilon_2 \wedge dx^6.
    \eea    
Note that in the above we have written the metric in the Einstein frame. From the form of the metric it is clear that the F-string
is along $x^6$ directions and delocalized in $x^1,\,x^2,\,\ldots, x^5$ directions whereas, the D5 brane is lying along $x^1,\,x^2,\,\ldots, x^5$
directions and delocalized in $x^6$ direction. $F^{(1)}_3$ and $F^{(2)}_3$ are the NSNS and RR 3-forms respectively. Again in the above
solution we have set the asymptotic values $\phi_0=0$ and $\chi_0=0$. The harmonic functions now have the forms $H_{1/5} = 1 + Q_{1/5}/r$, where
$Q_{1/5}$ are related to F-string and D5-brane charges as,
\be\label{chargereln}
e_1 = \frac{Q_1 \Omega_2 V_5}{\sqrt{2} \kappa_0}, \qquad g_5 = \frac{Q_5 \Omega_2 V_1}{\sqrt{2}\kappa_0}.
\ee
Here $\Omega_2=4\pi$ is the area of the unit two sphere and $V_1 = \int dx^6$, $V_5 = \int dx^1\wedge \ldots \wedge dx^5$.    
The electric-like (F, D1) string charge $e_1^{(i)}$ and the magnetic-like (NS5, D5) brane charge $g_5^{(i)}$ are given
in \eqref{e-charge} and \eqref{m-charge} respectively, however unlike in that case the integration region in this case
should be $R^5 \times S^2_\infty$ for the electric-like charge and $R^1 \times S^2_\infty$ for the magnetic-like charge. As before,
the electric-like and magnetic-like charges transform under SL(2,R) as $e_1 \to \Lambda e_1$ and $g_5 \to (\Lambda^{-1})^T g_5$,
with the standard choice of SL(2,R) matrix as given in \eqref{Lambda}.       
Since our initial classical configuration (F, D5) has classical F-string charge $\bar{\Delta}^{1/2}_1$ and classical D5-brane charge 
$\bar{\Delta}^{1/2}_5$, 
we have 
\be\label{initialstate}
e_1 = \left(\begin{array}{c}
\bar{\Delta}_1^{1/2} \\
0 \end{array}\right) e_{10},  \qquad g_5 = \left(\begin{array}{c}
0 \\
\bar{\Delta}_5^{1/2} \end{array}\right) g_{50},
\ee
where as usual $e_{10}$ and $g_{50}$ are the unit charges of the strings and the 5-branes. the SL(2,R) transformed charges after imposing
the quantization would be given as, $\bar{e}_1^T = (p,q)e_{10}$ and $\bar{g}_5^T = (m,n)g_{50}$, where $(p,q)$ and $(m,n)$ are two pairs
of integers. These quantization conditions give us the following relations  
\bea\label{fd5cossinalpha}
& & \cos\alpha = -(q\chi_0-p) e^{\phi_0/2} \bar{\Delta}_1^{-1/2} = (m\chi_0+n) e^{\phi_0/2} \bar{\Delta}_5^{-1/2},\nn
& & \sin\alpha = q\,e^{-\phi_0/2} \bar{\Delta}_1^{-1/2} = -m\, e^{-\phi_0/2} \bar{\Delta}_5^{-1/2} ,
\eea
and using these relations we determine
\bea\label{deltas}
& & \bar{\Delta}_1 = q^2 e^{-\phi_0} + (q\chi_0 -p)^2 e^{\phi_0}, \qquad \bar{\Delta}_5 = m^2 e^{-\phi_0} + (m\chi_0 + n)^2 e^{\phi_0},\nn
& & \tan\alpha = \frac{-q e^{-\phi_0}}{q\chi_0-p} = \frac{-m e^{-\phi_0}}{m\chi_0+n}.
\eea 
The last relation can be satisfied if $pm=-qn$. This in turn implies that the integers must have the forms $(p,q) = k(a,b)$
and $(m,n)=k'(-b,a)$, where $(a,b)$ and $(k,k')$ are two pairs of co-prime integers. Note that here the moduli $\phi_0$ and $\chi_0$ can 
be arbitrary unlike in the previous case. So this state is very similar to the 1/2 BPS ((F, D1), (NS5, D5)) one. The discussion of how to recover all 
the special case states from the general 1/4 BPS ((F, D1), (NS5, D5)) state in this case is also very similar to that given in \cite{Lu:1999uc} 
and therefore will not be repeated here. The complete SL(2,Z) invariant 1/4 BPS ((F, D1), (NS5, D5)) threshold bound state
solution where strings are perpendicular to the 5-branes is given below,
\bea\label{completefd5}
ds^2 &=&  - {\cal H}_1^{-3/4} {\cal H}_5^{-1/4} dt^2 + {\cal H}_1^{1/4} {\cal H}_5^{- 1/4} \sum^5_{i = 1} (d x^i)^2 \nn
          &\,& + {\cal H}_1^{-3/4} {\cal H}_5^{3/4} (dx^6)^2 + {\cal H}_1^{1/4} {\cal H}_5^{3/4} \left( dr^2 + r^2 d\Omega_2^2\right),\nn
    e^{2\phi} &=& \frac{\left(\gamma^2 {\cal H}_1 {\cal H}_5 + \d^2\right)^2}{{\cal H}_1 {\cal H}_5}, \quad \chi = 
\frac{\a\gamma {\cal H}_1 {\cal H}_5 + \b\d}{\gamma^2 {\cal H}_1 {\cal H}_5 + \d^2},\nn
    F^{(1)}_3 &=& \d\frac{Q_1}{{\cal H}_1^2 r^2} dt \wedge d x^6 \wedge dr - \gamma Q_5 \epsilon_2 \wedge dx^6,\nn
    F^{(2)}_3 &=& -\b \frac{Q_1}{{\cal H}_1^2 r^2} dt \wedge dx^6 \wedge dr + \a Q_5 \epsilon_2 \wedge dx^6,
    \eea    
where ${\cal H}_{1/5} = 1 + Q_{1/5}/r$ are the new harmonic functions with the same form as the old ones, but $Q_{1/5}$ have now changed to
$\bar{\Delta}_{1/5}^{1/2} Q_0$. Here $\bar{\Delta}_{1/5}$ are as given in \eqref{deltas} and $Q_0 = \sqrt{2}\kappa_0 Q_0^5/(\omega_2V_1)$,
with $Q_0^5$ as defined before. The SL(2,R) parameters $\a,\,\b,\,\gamma,\,\d$ can be obtained from \eqref{Lambda} with the use of
\eqref{fd5cossinalpha} and are given as,
\bea\label{fd5abcd}
& & \alpha = p \bar{\Delta}_1^{-\half} =  n \bar{\Delta}_5^{-\half},\nn 
& &  \beta =  -\left[qe^{-\phi_0} + \chi_0(q\chi_0 -p)e^{\phi_0}\right]\bar{\Delta}_1^{-\half} =  
\left[me^{-\phi_0} + \chi_0(m\chi_0+n)e^{\phi_0}\right]\bar{\Delta}_5^{-\half},\nn
& & \gamma =  q \bar{\Delta}_1^{-\half} = -m\bar{\Delta}_5^{-\half},\nn
& & \delta = -(q\chi_0 - p) e^{\phi_0} \bar{\Delta}_1^{-\half} = (m\chi_0+n)e^{\phi_0}\bar{\Delta}_5^{-\half}.
\eea
As before we notice here that the SL(2,R) transformed metric in \eqref{completefd5} retains its form as the original metric given
in \eqref{fd5} (as the metric is given in the Einstein frame) except for $Q_{1/5}$ which have been changed due to the corresponding charge quantizations, respectively. But the
dilaton has been changed and a $\chi$ has been generated due to the SL(2,R) transformation even though the original solution does not contain 
a $\chi$. The 3-form fields have also been changed but no 5-form field has been generated as in the previous case and unlike the
1/2 BPS case. Again it can be easily checked that as $r\to \infty$, ${\cal H}_{1/5} \to 1$ and therefore, $\phi \to \phi_0$ and
$\chi \to \chi_0$ as expected.

One can calculate the string frame tension of this bound state from the ADM mass per unit 5-brane volume following \cite{Lu:1999uca} 
and \cite{Lu:1993vt} and we get,
\be\label{tensionfd5}
T_5(k,k';a,b) = \frac{T_0^5}{g} \left(k+\frac{k'}{g}\right)\sqrt{b^2 + (b\chi_0 - a)^2 g^2}
\ee
where $T_0^5$ is as defined before and $g=e^{\phi_0}$ is the string coupling. This tension is the sum of the tensions of  
$(p,q)$ strings and $(m,n)$ 5-branes. One can easily check that this expression gives all the expected tensions of
the constituent branes.

Using the Table \ref{t1} one can generate various 1/4 BPS descendant states, by applying T-duality, starting 
from this 1/4 BPS ((F, D1), (NS5, D5)) state. So, for example, applying T-duality along $x^6$, we can get 1/4 BPS ((W, D0), (KK, D6)) 
state and applying T-duality along $x^5$, we can get 1/4 BPS ((F, D2), (NS5, D4)) and again applying T-duality along $x^4$ on this
state we can get 1/4 BPS ((F, D3), (NS5, D3)) state and so on. All these states are asymptotically flat and other such asymptotically flat states
cannot be generated by applying T-duality along the directions transverse to both strings and 5-branes in this case.     

\section{Conclusion}

On one hand it is known that when (F, D1) or $(p,q)$ strings are placed parallel to one of the world-volume spatial directions
of (NS5, D5) or $(m,n)$ 5-branes, they should attract each other since F-strings attract D5 branes and D1-strings attract NS5 branes
and therefore, when they form bound state, it should only be 1/2 BPS ((F, D1), (NS5, D5)) state. On the other hand, we know
that type IIB string theory admits 1/4 BPS (D1, D5) bound state solution and so by applying the SL(2,R) classical symmetry of
type IIB string theory and imposing charge quantization we should be able to construct an SL(2,Z) invariant 1/4 BPS ((F, D1), (NS5, D5))
bound state in this theory where $(p,q)$ strings and $(m,n)$ 5-branes should not feel any attractive force on each other. In this
paper we have resolved this apparent contradiction by explicitly constructing this new vacua, i.e., 1/4 BPS ((F, D1), (NS5, D5)) 
state of type IIB string theory from 1/4 BPS (D1, D5) state and applying SL(2,R) symmetry as well as the charge quantization condition to it.
We found that the consistency condition for the existence of this bound state manifests itself in the form of certain constraint
given by \eqref{modulicondition} (or \eqref{keymodulicondition} or \eqref{keyonemodulicondition}) on the vaccum moduli of the 
solution. We have discussed and pointed out many subtleties on how to recover the special case bound states from this general bound
state solution not faced before in similar constructions. We then computed the force experienced by a probe (F, D1) or $(p,q)$ string
placed in the background of (NS5, D5) or $(m,n)$ 5-branes. We found that when the string is  parallel to one of the spatial
world-volume directions of the 5-brane, the force can vanish under certain condition. Nicely, we found that
the no-force condition precisely matches with the condition on the moduli \eqref{modulicondition} we obtained before while
constructing the SL(2,Z) invariant 1/4 BPS ((F, D1), (NS5, D5)) state. Further we have given the tension formula for this
general bound state from the ADM mass and also discussed how to obtain various other descendant 1/4 BPS bound states from this
by the application of T-dualities in various isometric directions.

For completeness, we have also constructed another SL(2,Z) invariant new vacua of type IIB string theory in the form of a 1/4 BPS 
((F, D1), (NS5, D5)) bound state where, strings are perpendicular to 5-branes. In this case we started from the known 1/4 BPS
(F, D5) bound state solution of type IIB theory and applied SL(2,R) transformation as well as the charge quantization condition to it. 
We pointed out that the construction of this
state is very similar to that of the 1/2 BPS ((F, D1), (NS5, D5)) state, constructed before by two of us in \cite{Lu:1999uc}.
In this case there is no extra condition on the vacuum moduli unlike in the previous case and they can be completely arbitrary.
Here also we have given the tension formula and discussed how the other descendant 1/4 BPS bound states can be obtained by T-duality.     

\section*{Acknowledgements}
We would like to thank our anonymous referees for fruitful suggestions which help us to improve the manuscript and also for bringing reference 13 to our attention.  QJ and JXL acknowledge support by a key grant from the NSF of China with Grant No: 11235010.

\end{document}